\documentclass[twocolumn,showpacs,preprintnumbers]{revtex4}
\usepackage{amssymb}
\usepackage{amsfonts}
\usepackage{amsmath}
\usepackage{graphicx}
\usepackage{dcolumn}
\usepackage{bm}

\input{tcilatex}
\begin{document}

\preprint{APS/123-QED}
\title{Asymmetric Mach-Zehnder atom interferometers}
\author{B. Dubetsky}
\email{bdubetsky@gmail.com}
\affiliation{US Army Research Laboratory, Adelphi, MD 20783}
\affiliation{Department of Physics, Stanford University, Stanford, California 94305, USA}
\date{\today }

\begin{abstract}
It is shown that using beam splitters with non-equal wave vectors results in
a new recoil diagram which is qualitatively different from the well-known
diagram associated with the Mach-Zehnder atom interferometer. We predict a
new asymmetric Mach-Zehnder atom interferometer (AMZAI) and study it when
one uses a Raman beam splitter. The main feature is that the phase of AMZAI
contains a quantum part proportional to the recoil frequency. A response
sensitive only to the quantum phase was found.{\large \ }A new technique to
measure the recoil frequency and fine structure constant is proposed and
studied outside of the Raman-Nath approximation.
\end{abstract}

\pacs{03.75.Dg; 37.25.+k; 04.80.-y}
\maketitle

It is well-known that atom interferometry \cite{c1} is caused by the
quantization of the atomic center-of-mass motion. When the incident atomic
momentum state $\left\vert \mathbf{p}\right\rangle $ splits into two states $%
\left\vert \mathbf{p}\right\rangle $ and $\left\vert \mathbf{p}+\hbar 
\mathbf{k}\right\rangle $ after passing through a beam splitter having
effective wave vector $\mathbf{k},$ the coherence between these states
evolves as 
\begin{equation}
\rho \left( \mathbf{p}+\hbar \mathbf{k},\mathbf{p},t\right) \propto \exp
\left( -i\omega _{\mathbf{p}+\hbar \mathbf{k},\mathbf{p}}t\right)  \label{1}
\end{equation}%
where the frequency of transition between states%
\begin{equation}
\omega _{\mathbf{p}+\hbar \mathbf{k},\mathbf{p}}=\dfrac{1}{2M\hbar }\left[
\left( \mathbf{p}+\hbar \mathbf{k}\right) ^{2}-\mathbf{p}^{2}\right] =%
\mathbf{k}\cdot \dfrac{\mathbf{p}}{M}+\dfrac{\hbar \mathbf{k}^{2}}{2M}
\label{2}
\end{equation}%
contains the quantum term, recoil frequency%
\begin{equation}
\omega _{k}=\dfrac{\hbar \mathbf{k}^{2}}{2M},  \label{3}
\end{equation}%
where $M$ is an atomic mass. When $t$ is of the order of interrogation time $%
T$ of the given atomic interferometer, one can expect the phase of the
interferometer to contain the quantum contribution 
\begin{equation}
\phi _{q}\thicksim \omega _{k}T,  \label{4}
\end{equation}%
which would reveal the quantum nature of the atom interference. This phase
leads to the recoil splitting of the optical Ramsey fringes \cite{c16},
Talbot effect, i.e. quantum beats of the atom interferometer signal, in the
standing wave fields \cite{c1,c17,c20} and microfabricated structures \cite%
{c18}. Precise measurement of the phase (\ref{4}) was performed \cite{c9}
using a Raman analogue of the atom interferometer \cite{c19} involving 4
time-separated counterpropagating traveling waves. The value of $\omega _{k}$
was proposed for measuring the fine structure constant \cite{c9}, which
resulted \cite{c14} in a resolution in this constant of 0.25 ppb.

In spite of this, the phase of the well-known and widely used Mach-Zehnder
atom interferometer (MZAI), in which an atom passes through 3 beam splitters
separated in time with a delay $T$, contains no quantum term. In the uniform
gravitational field $\mathbf{g}$ the phase is given by \cite{c2}%
\begin{equation}
\phi =\mathbf{k}\cdot \mathbf{g}T^{2}.  \label{5}
\end{equation}%
The reason is that quantum corrections affect the atomic position at the
moments of interaction with the 2nd and 3rd beam splitters, but in the
uniform gravity field these contributions cancel one another (see for
example Appendix in \cite{c3}). We would like to underline that derivations\
of the MZAI phase is purely quantum (see examples of this derivation in \cite%
{c2,c3,c4}),but the result of those derivations (\ref{5}) is purely
classical. The quantum contribution to the MZAI phase arises in the rotating
frame \cite{c4.1}, or in the non-uniform field, in the presence of the
gravity-gradient tensor \cite{c4.2,c8} or in the presence of the gravity
curvature tensor \cite{c3}, or in the strongly non-uniform field of the
external test mass \cite{c3}, the quantum part of the phase in
gravity-gradient field of the external test mass has been recently observed 
\cite{c33}.

The absence of the quantum phase (\ref{4}) allows one to doubt that MZAI is
caused by matter wave interference. In this article, I propose a
modification of the MZAI which contains the term (\ref{4}). Moreover, we
found a response that contains only the quantum phase. Since it is
insensitive to gravity, vibration, phase and frequency noise, this response
can be used to measure recoil frequency and fine structure constant. An
advantage of our approach compared to the approach, using conjugate atom
interferometers in the Bragg regime \cite{c3.1}, is that our effect is
insensitive to phase and frequency noise and has no diffraction phase \cite%
{c3.2}.

In our modification, we first tried to use beam splitters with\ different
effective wave vectors. If the beam splitter is a pulse of standing wave 
\cite{c1}, or Raman pulse \cite{c4}, then usually the field consists of two
counterpropagating traveling waves having\ wave vectors $\mathbf{q}_{1}$ and 
$\mathbf{q}_{2}\approx -\mathbf{q}_{1}.$ The effective wave vector of this
beam splitter is%
\begin{equation}
\mathbf{k}=\mathbf{q}_{1}-\mathbf{q}_{2}\approx 2\mathbf{q}_{1}.  \label{6}
\end{equation}%
One can obtain a value of $\mathbf{k}$ different from that given by Eq. (\ref%
{6}) if the traveling waves are not counterpropagating $\left( \mathbf{q}%
_{1}\not=-\mathbf{q}_{2}\right) .$ In this case, the magnitude of the
effective wave vector $\left\vert \mathbf{k}\right\vert =k\sin \alpha ,$
where $\alpha $ is the half-angle between vectors $\mathbf{q}_{1}$ and $%
\mathbf{q}_{2}$ and $k\approx 2q_{1}.$ Such a field was used for the first
observation of atomic focusing \cite{c5}, in the theory of conical lens for
atoms \cite{c6}, to consider quasiperiodic Fresnel atom optics \cite{c7}, to
separate in time double-loop atomic gyro and stimulated echo \cite{c8}.
Another way of getting non-equal wave vectors is to use the sequential
technique \cite{c9,c10} to increase $\mathbf{k}$ and apply it with different
amounts of additional $\pi -$pulses near the 1st, 2nd and 3rd pulses. The
sequential technique \cite{c10} is more promising than technique combining
Bragg diffraction and Bloch oscillations \cite{c14}. We are expecting that
AMZAI effective wave \ vectors can be increased using the sequential
technique and do not consider the Bloch oscillations in this article. MZAI
with a slightly different effective wave vector of the second Raman pulse
has been also proposed \cite{c10.1} to vanish sensitivity to the atom
clouds' initial position and the velocity caused by the gravity-gradient
tensor. This effect has been recently observed \cite{c10.2,c10.3}. The
difference from our case is that (a) we allow an arbitrary change of the
wave vectors' magnitudes, (b) but we do not violate the phase-matching
conditions, (c) the gravity field is uniform (on this stage we neglect the
gravity gradient terms), (d) the asymmetry proposed in \cite{c10.1} does not
lead and is not aimed at leading to any quantum term in the MZAI phase.

The recoil diagram of the MZAI with non-equal wave vectors is shown in Fig. %
\ref{f1}. Calculations show that the phase of this interferometer still has
no quantum part (see appendix \ref{a}).

\begin{figure}[!t]
\includegraphics[width=8cm]{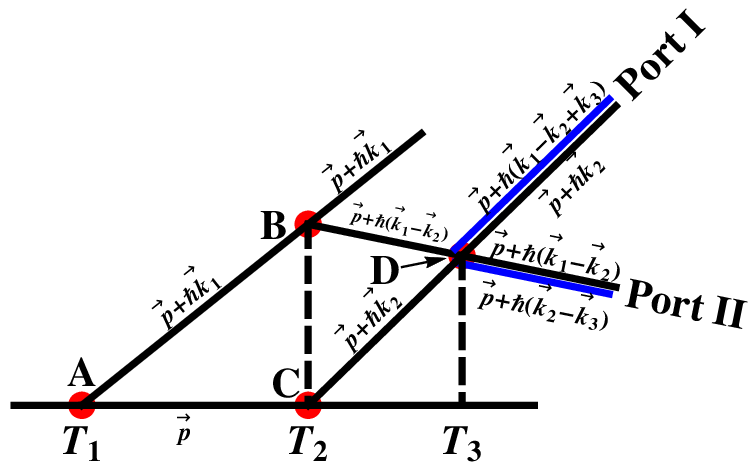}
\caption{MZAI with different effective wave vectors. If $\left\vert \mathbf{p%
}\right\rangle $ is an incident momentum state, then the beam splitter
produces, in addition to state $\left\vert \mathbf{p}\right\rangle ,$ the
scattered momentum state $\left\vert \mathbf{p}\pm \hbar \mathbf{k}%
_{i}\right\rangle $. The scattered momentum states after the 3rd beam
splitter are shown in blue.}
\label{f1}
\end{figure}

\begin{figure}[!t]
\includegraphics[width=8cm]{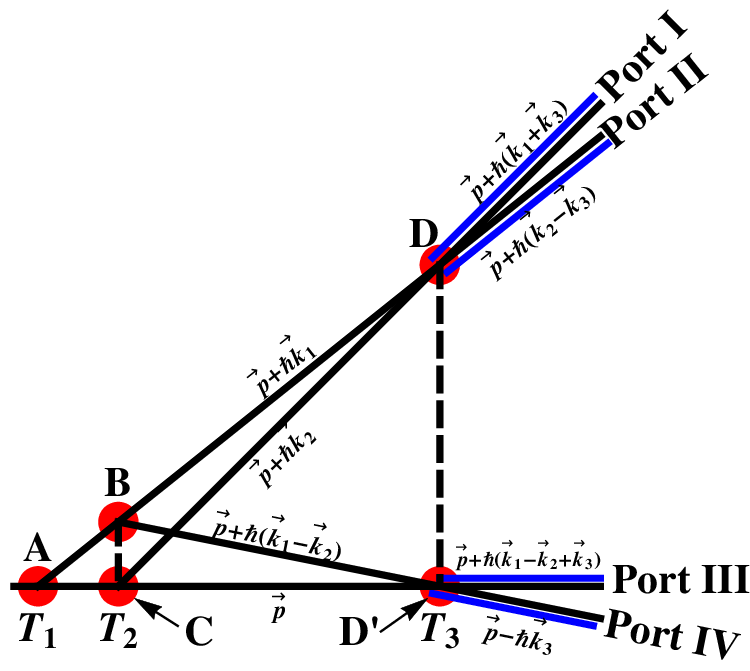}
\caption{Recoil diagram of the AMZAI.}
\label{f2}
\end{figure}

Nevertheless, one notices that lines $AB$ and $CD$ are no longer parallel,
and at some moment in time they cross one another. Applying at that time the
3rd beam splitter, one obtains new an asymmetric Mach-Zehnder atom
interferometer (AMZAI). One can say the same thing about lines $AC$ and $BD.$
The corresponding recoil diagram for this interferometer is shown in Fig. %
\ref{f2}.

It is evident from the figure that the choice of the wave vectors and times $%
T_{i}$\ has to obey 2 constraints. The first constraint is that the blue and
black lines in each port have to be parallel, i.e.%
\begin{equation}
\mathbf{k}_{1}=\mathbf{k}_{2}-\mathbf{k}_{3},  \label{6.2}
\end{equation}%
which is a well-known phase matching condition. The second constraint is
that $D$\ is the point of crossing and therefore $\overrightarrow{AD}=%
\overrightarrow{AC}+\overrightarrow{CD},$\ which means that%
\begin{equation}
\mathbf{k}_{1}\left( T_{3}-T_{1}\right) =\mathbf{k}_{2}\left(
T_{3}-T_{2}\right) .  \label{6.3}
\end{equation}%
One can get the same equality from the constraint $\overrightarrow{%
AD^{\prime }}=\overrightarrow{AB}+\overrightarrow{BD^{\prime }}.$ Choosing
the wave vector $\mathbf{k}_{2}$ as an independent variable, one can get
from constraints (\ref{6.2},\ref{6.3}) that other wave vectors have to be
equal to\ 
\begin{subequations}
\label{13}
\begin{eqnarray}
\mathbf{k}_{1} &=&\left( 1-s\right) \mathbf{k}_{2},  \label{13a} \\
\mathbf{k}_{3} &=&s\mathbf{k}_{2},  \label{13b}
\end{eqnarray}%
where 
\end{subequations}
\begin{equation}
s=\left( T_{2}-T_{1}\right) /\left( T_{3}-T_{1}\right) .  \label{13.1}
\end{equation}

\begin{figure}[!t]
\includegraphics[width=8cm]{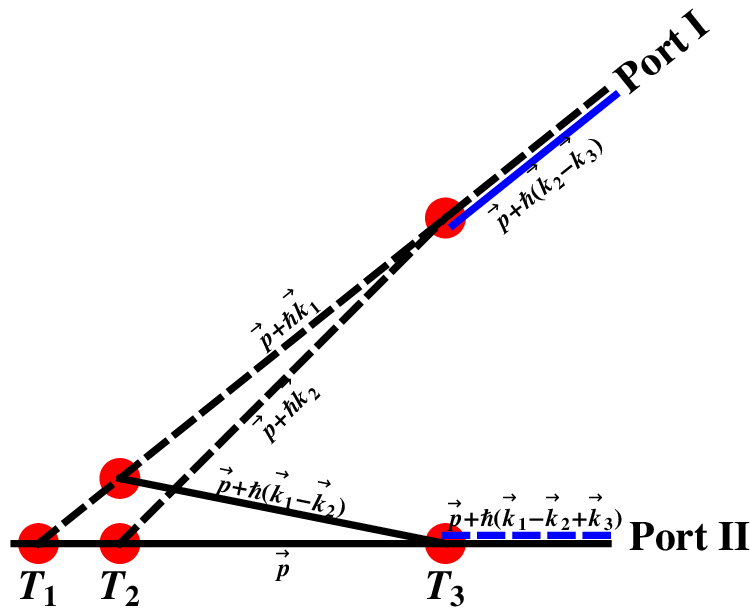}
\caption{AMZAI in the case of Raman beam splitter for colinear beam spliters 
$\mathbf{k}_{1}\Vert \mathbf{k}_{2}\Vert \mathbf{k}_{3}$ and $k_{2}>k_{1}.$
Solid and dashed lines correspond to the atomic ground and excited states.}
\label{f3}
\end{figure}

If one uses a Raman pulse for beam splitting, then the change of momentum is
accompanied by a change of the atomic internal states $\left\vert
g\right\rangle $ (ground) and $\left\vert e\right\rangle $ (excited).
Specifically, the incident state $\left\vert g,\mathbf{p}\right\rangle $
splits into states $\left\vert g,\mathbf{p}\right\rangle $ and $\left\vert e,%
\mathbf{p}+\hbar \mathbf{k}\right\rangle ,$ while the incident state $%
\left\vert e,\mathbf{p}\right\rangle $ splits into states $\left\vert e,%
\mathbf{p}\right\rangle $ and $\left\vert g,\mathbf{p}-\hbar \mathbf{k}%
\right\rangle $. From the recoil diagram corresponding to these rules shown
in Fig. \ref{f3}, one finds in each of the output ports different atomic
internal states which cannot interfere with each other. 

A similar
situation occurs for a Stern-Gerlach beam splitter, where the use of an
additional microwave pulse changes the internal atomic label and leads to
interference \cite{c11}. I also propose applying here the microwave pulse
resonant to transition $g\rightarrow e$ just before (the case considered
here) or after 3rd Raman pulse \cite{c11.1}. It is easier to perform
calculations than to draw the recoil diagram corresponding to this set of
pulses.

Let's assume that atoms are launched at $t=0$ and interact with 3 Raman
pulses. Pulse $i$ is applied at time $T_{i}$ and has the area, effective
wave vector, phase and Raman detuning $\theta _{i},\mathbf{k}_{i},\phi _{i}$
and $\delta _{i}$. In addition the microwave pulse, having the area,
effective wave vector, phase and detuning $\theta _{m},\mathbf{k}_{m}=0,\phi
_{m}$ and $\delta _{m}$, is applied at time $T_{m}.$ I assume, for
simplicity, that all pulses have the same duration $\tau .${\Huge \ }The
calculations were performed using Wigner representation for the atomic
density matrix. To get the AMZAI phase, I used equations for the density
matrix evolution in times between pulses and equations for the density
matrix change immediately after the interaction with a given pulse, i.e.
Eqs. (21, 48) in \cite{c8}. Regarding the microwave pulse time $T_{m},$ I
assume that 
\begin{equation}
0<T_{3}-T_{m}\ll T_{3}  \label{7}
\end{equation}%
so that one can ignore the density matrix evolution at $T_{m}+\tau <t<T_{3}$%
. If initially all atoms are in the ground state,%
\begin{equation}
\rho _{gg}\left( \mathbf{x},\mathbf{p},0\right) =f\left( \mathbf{x},\mathbf{p%
}\right) ,\rho _{ee}\left( \mathbf{x},\mathbf{p},0\right) =\rho _{eg}\left( 
\mathbf{x},\mathbf{p},0\right) =0,  \label{8}
\end{equation}%
then I calculate the probability of atoms' excitation immediately at time $%
T_{3}+\tau $%
\begin{equation}
w=\int d\mathbf{p}d\mathbf{x}\rho _{ee}\left( \mathbf{x},\mathbf{p}%
,T_{3}+\tau \right) .  \label{9}
\end{equation}%
The straightforward, but extended, calculations of response (\ref{9}) will
be omitted here. The interaction which I consider here is a particular case
of the interaction with 4 pulses considered previously in Sec. VII of the
article \cite{c8}, where the 3rd pulse is a microwave pulse, the 4th pulse
becomes the 3rd one, and the processes shown in Fig. \ref{f2} correspond to
processes called in \cite{c8} "stimulated echo". Saving only these processes
and background terms one finds%
\onecolumngrid
\begin{subequations}
\label{10}
\begin{eqnarray}
w &=&\bar{w}+w_{I+}+w_{I-}+w_{II+}+w_{II-},  \label{10a} \\
\bar{w} &=&\dfrac{1}{2}\left( 1-\cos \theta _{1}\cos \theta _{2}\cos \theta
_{m}\cos \theta _{3}\right) ,  \label{10b} \\
\left\{ 
\begin{array}{c}
w_{I\pm } \\ 
w_{II\pm }%
\end{array}%
\right\} &=&-\dfrac{1}{8}\sin \theta _{3}\sin \theta _{m}\sin \theta
_{2}\sin \theta _{1}\int d\mathbf{x}d\mathbf{p}f\left( \mathbf{x},\mathbf{p}%
\right) \left\{ 
\begin{array}{c}
\cos \left( \phi _{I\pm }+\phi _{I}\right) \\ 
\cos \left( \phi _{II\pm }+\phi _{II}\right)%
\end{array}%
\right\} ,  \label{10c} \\
\phi _{I,II} &=&-\delta _{3}T_{3}+\delta _{m}T_{m}\pm \delta _{2}T_{2}\mp
\delta _{1}T_{1}-\phi _{3}+\phi _{m}\pm \phi _{2}\mp \phi _{1}-\arg \left(
\Omega _{3}^{\ast }\Omega _{m}\right) \mp \arg \left( \Omega _{2}\Omega
_{1}^{\ast }\right) ,  \label{10c.1} \\
\left\{ 
\begin{array}{c}
\phi _{I\pm } \\ 
\phi _{II\pm }%
\end{array}%
\right\} &=&\left\{ 
\begin{array}{c}
\mathbf{k}_{3}\cdot \mathbf{x}_{3\pm }-\mathbf{k}_{2}\cdot \mathbf{x}_{2}+%
\mathbf{k}_{1}\cdot \mathbf{x}_{1} \\ 
\mathbf{k}_{3}\cdot \mathbf{x}_{3\pm }+\mathbf{k}_{2}\cdot \mathbf{x}_{2}-%
\mathbf{k}_{1}\cdot \mathbf{x}_{1}%
\end{array}%
\right\} ,  \label{10d} \\
\left\{ \mathbf{x}_{1},\mathbf{p}_{1}\right\} &=&\left\{ \mathbf{X}\left( 
\mathbf{x},\mathbf{p},T_{1}\right) ,\mathbf{P}\left( \mathbf{x},\mathbf{p}%
,T_{1}\right) +\dfrac{\hbar \mathbf{k}_{1}}{2}\right\} ,  \label{10e} \\
\left\{ \mathbf{x}_{2},\mathbf{p}_{2\pm }\right\} &=&\left\{ \mathbf{X}%
\left( \mathbf{x}_{1},\mathbf{p}_{1},T_{2}-T_{1}\right) ,\mathbf{P}\left( 
\mathbf{x}_{1},\mathbf{p}_{1},T_{2}-T_{1}\right) \mp \dfrac{\hbar \mathbf{k}%
_{2}}{2}\right\} ,  \label{10f} \\
\mathbf{x}_{3\pm } &=&\mathbf{X}\left( \mathbf{x}_{2},\mathbf{p}_{2\pm
},T_{3}-T_{2}\right) ,  \label{10g}
\end{eqnarray}%
where $\Omega _{n}$\ is a two-photon Raman Rabi frequency of the Raman pulse 
$n,$ $\Omega _{m}$\ is a Rabi frequency of the microwave pulse$,$ $\left\{ 
\mathbf{X}\left( \mathbf{x},\mathbf{p},t\right) ,\mathbf{P}\left( \mathbf{x},%
\mathbf{p},t\right) \right\} $ is a point of the atom trajectory in the
phase space, and $\left\{ \mathbf{x},\mathbf{p}\right\} $ is the initial
point of this trajectory. I consider here for simplicity only atom motion in
the uniform gravity field $\mathbf{g}$, when 
\end{subequations}
\begin{subequations}
\label{11}
\begin{eqnarray}
\mathbf{X}\left( \mathbf{x},\mathbf{p},t\right) &=&\mathbf{x}+\dfrac{\mathbf{%
p}}{M}t+\mathbf{g}\dfrac{t^{2}}{2},  \label{11a} \\
\mathbf{P}\left( \mathbf{x},\mathbf{p},t\right) &=&\mathbf{p}+M\mathbf{g}t.
\label{11b}
\end{eqnarray}%
Using these dependencies one obtains 
\end{subequations}
\begin{eqnarray}
\phi _{I\pm } &=&\left( \mathbf{k}_{3}-\mathbf{k}_{2}+\mathbf{k}_{1}\right)
\cdot \mathbf{x}+\left( \mathbf{k}_{3}T_{3}-\mathbf{k}_{2}T_{2}+\mathbf{k}%
_{1}T_{1}\right) \cdot \dfrac{\mathbf{p}}{M}+\dfrac{1}{2}\left( \mathbf{k}%
_{3}T_{3}^{2}-\mathbf{k}_{2}T_{2}^{2}+\mathbf{k}_{1}T_{1}^{2}\right) \cdot 
\mathbf{g}  \notag \\
&&+\dfrac{\hbar }{2M}\left\{ \mathbf{k}_{3}\cdot \left[ \mathbf{k}_{1}\left(
T_{2}-T_{1}\right) +\left( \mathbf{k}_{1}\mp \mathbf{k}_{2}\right) \left(
T_{3}-T_{2}\right) \right] -\mathbf{k}_{2}\cdot \mathbf{k}_{1}\left(
T_{2}-T_{1}\right) \right\}  \label{12}
\end{eqnarray}%
\twocolumngrid%
One sees that, owing to the first two terms in Eq. (\ref{12}), the integrand
for $w_{I\pm }$ in Eq. (\ref{10c}) rapidly oscillates in the phase space,
which can wash out the interference signal.. However, owing to the
constraints (\ref{6.2}, \ref{6.3}) the first two terms are exactly equal to
0. Substituting Eq. (\ref{13}) into Eq. (\ref{12}), one finds 
\begin{subequations}
\label{14}
\begin{eqnarray}
\phi _{I\pm } &=&\phi _{c}\mp \phi _{q},  \label{14a} \\
\phi _{c} &=&\dfrac{1}{2}\mathbf{k}_{2}\cdot \mathbf{g}\left(
T_{3}-T_{1}\right) ^{2}s\left( 1-s\right) ,  \label{14b} \\
\phi _{q} &=&\omega _{k_{2}}\left( T_{3}-T_{1}\right) s\left( 1-s\right) .
\label{14c}
\end{eqnarray}%
One can find that for wave vectors (\ref{13}), terms $w_{II\pm }$ are washed
out and the sum of the interference terms $w_{I+}+w_{I-}$ contains
separately a factor depending on the quantum phase (\ref{14c}) and a factor
depending on the all the other phases, $\phi _{c},~\phi _{i},~\delta
_{i}T_{j},$ i.e. 
\end{subequations}
\begin{equation}
w=\bar{w}-\dfrac{1}{4}\sin \theta _{3}\sin \theta _{m}\sin \theta _{2}\sin
\theta _{1}\cos \left( \phi _{q}\right) \cos \left( \phi _{c}+\phi
_{I}\right) ,  \label{14.1}
\end{equation}

If one monitors the signal (\ref{14.1}) as a function of any phase or
detuning of the Raman or microwave field, then the difference of the maximum
and minimum of the signal is given by 
\begin{equation}
A=w_{\max }-w_{\min }=\dfrac{1}{2}\left\vert \sin \theta _{3}\sin \theta
_{m}\sin \theta _{2}\sin \theta _{1}\cos \left( \phi _{q}\right) \right\vert
.  \label{15}
\end{equation}%
\textbf{The oscillating dependence in (\ref{15}) on the interrogation time }$%
T_{3}-T_{1}$\textbf{\ is caused only by the quantization of the atomic
center-of-mass motion and reveals the quantum nature of the atom
interference. }Evidently, the signal (\ref{15}) is insensitive to the
gravity field, vibration noise, and phase noise of the laser fields.

The quantum phase is maximal when $k_{2}=k$ and $s=1/2,$ or $%
k_{1}=k_{3}=k/2,~T_{2}=\left( T_{1}+T_{3}\right) /2.$ This means that the
2nd pulse still consists of counterpropagating fields, but the 1st and 3rd
pulses have to consist of traveling waves having $\pi /3$ angle between
their wave vectors. In this case%
\begin{equation}
\phi _{q}=\dfrac{1}{4}\omega _{k}\left( T_{3}-T_{1}\right) .  \label{16}
\end{equation}%
The contrast of the signal (\ref{15}), $c=\left. \left( w_{\max }-w_{\min
}\right) _{\phi _{q}=0}\right/ \bar{w}$ achieves 100\% for $\pi /2-\pi
/2-\pi /2-\pi /2$ sequence of the pulses.

Phases $\phi _{II\pm }$ can be considered in the same manner. To avoid
washing out the $w_{II\pm }$ terms, one has to choose the same value (\ref%
{13a}) of the wave vector $\mathbf{k}_{1}$ and opposite value of the wave
vector $\mathbf{k}_{3},$ $\mathbf{k}_{3}=-s\mathbf{k}_{2}.$With this choice
of wave vectors, the terms $w_{I\pm }$ are washed out, while the phases
change their sign, $\phi _{II\pm }=-\phi _{I\pm }.$

The equations for the density matrix in the Wigner representation, derived
in \cite{c8} have a limited region of validity. One can use them only if the
Raman detunings $\delta _{n}$\ compensate for the recoil frequency $\omega
_{k_{n}}$. One can easily achieve the compensation for MZAI, since all beam
splitters have the same effective wave vector $\mathbf{k}.$\ The situation
changes for AMZAI: since one could not achieve compensation for all wave
vectors simultaneously, the approach \cite{c8} becomes valid only in the
Raman-Nath approximation, when the pulse duration $\tau $\ is significantly
smaller than an inverse recoil frequency{\LARGE \ }%
\begin{equation}
\omega _{k_{n}}\tau \ll 1.  \label{19}
\end{equation}%
For the 2-quantum transition in Rb$^{87}$\ $\omega _{k}^{-1}\approx 10.6\mu $%
s. The minimal duration of the pulses used in the atom interferometry is $50$%
ns \cite{c21}, which is sufficient for the Raman-Nath approximation. But in
experiments exploring Raman beam splitters, pulse durations are in the range 
$\left( 4\div 35\right) \mu $s \cite%
{c10,c22,c23,c24,c25,c26,c27,c28,c29,c30,c31,c32}. To describe AMZAI outside
of the Raman-Nath approximation, I used the equation for an atomic wave
function in momentum space. The solution of this equation in the uniform
field \cite{c11} can be applied for both the wave function evolution between
pulses and inside a given Raman pulse if one chirps the pulse frequency with
a rate $\alpha _{n}$\ and{\LARGE \ }%
\begin{equation}
\left\vert \alpha _{n}-\mathbf{k}_{n}\cdot \mathbf{g}\right\vert \tau
^{2}\ll 1,  \label{20}
\end{equation}%
(see appendices \ref{b} and \ref{c}). The background of the probability of
excitation $\bar{w}$ is given by Eq. (\ref{a30}), while for the
interferometric part of this probability one obtains from Eqs.(\ref{a28a}, %
\ref{a29a})

\onecolumngrid%
\begin{subequations}
\label{21}
\begin{eqnarray}
\tilde{w} &=&-2\left\vert x\right\vert \cos \left( \phi _{g}\right) \sqrt{%
\left\vert x_{+}\right\vert ^{2}+\left\vert x_{-}\right\vert
^{2}+2\left\vert x_{+}\right\vert \left\vert x_{-}\right\vert \cos \left[
2r+\arctan \dfrac{\func{Im}\left( x_{-}\right) }{\func{Re}\left(
x_{-}\right) }-\arctan \dfrac{\func{Im}\left( x_{+}\right) }{\func{Re}\left(
x_{+}\right) }\right] },  \label{21a} \\
\phi _{g} &=&\phi _{I}+\phi _{c}+\left( \delta _{m}-\delta _{1,\mathbf{P}%
_{0}}+\omega _{k_{2}}\left( 1-s\right) ^{2}\right) \tau  \notag \\
&&-\arctan \dfrac{\func{Im}\left( x\right) }{\func{Re}\left( x\right) }%
-\arctan \dfrac{\cos r\func{Im}\left( x_{+}+x_{-}\right) -\sin r\func{Re}%
\left( x_{+}-x_{-}\right) }{\cos r\func{Re}\left( x_{+}+x_{-}\right) +\sin r%
\func{Im}\left( x_{+}-x_{-}\right) },  \label{21b}
\end{eqnarray}%
{\LARGE \ }where $\mathbf{P}_{0}$ is\ an initial atomic momentum and $\delta
_{n,\mathbf{P}}$ is a detuning (\ref{a19b}) associated with pulse $n.$
Expression (\ref{21}) has been derived for a rectangular shape of the Raman
and microwave pulses. One can accept that the pulses are rectangular if the
durations of the forward and backward fronts are significantly smaller than
the inverse recoil frequency.

The quantum phase $\phi _{q}$ is included now in parameter $r$ [see Eq. (\ref%
{a29b})]. It cannot now be separated from other phase factors. In spite of
this, after monitoring the signal (\ref{21a}), as a function of any Raman
field phase, one finds that the response (\ref{15}) still depends only on
the quantum phase, 
\end{subequations}
\begin{equation}
A=4\left\vert x\right\vert \sqrt{\left\vert x_{+}\right\vert ^{2}+\left\vert
x_{-}\right\vert ^{2}+2\left\vert x_{+}\right\vert \left\vert
x_{-}\right\vert \cos \left[ 2r+\arctan \dfrac{\func{Im}\left( x_{-}\right) 
}{\func{Re}\left( x_{-}\right) }-\arctan \dfrac{\func{Im}\left( x_{+}\right) 
}{\func{Re}\left( x_{+}\right) }\right] },  \label{22}
\end{equation}%
\twocolumngrid
One can define the magnitude of the response (\ref{22}) as the difference $%
R=A_{\max }-A_{\min },$\ which is equal to%
\begin{equation}
R=8\left\vert x\right\vert Min\left\{ \left\vert x_{+}\right\vert
,\left\vert x_{-}\right\vert \right\} .  \label{23}
\end{equation}

\begin{figure}[!t]
\includegraphics[width=8cm]{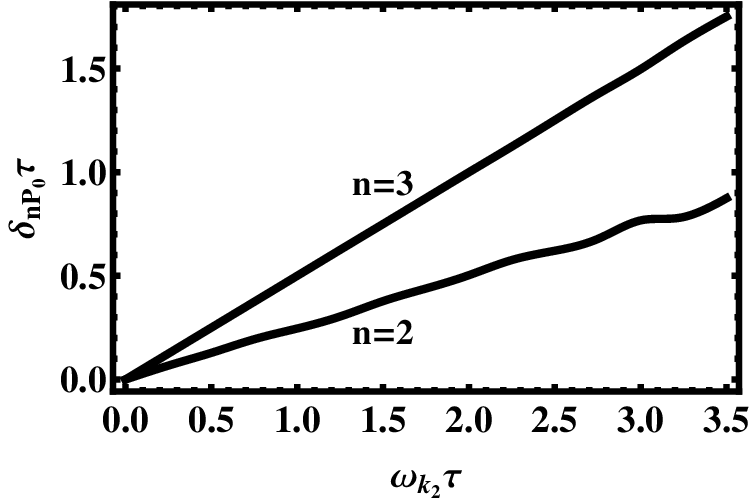}
\caption{Dependences of the optimum values of the Raman detunings on the
atomic recoil frequency.}
\label{f4}
\end{figure}

\begin{figure}[!t]
\includegraphics[width=8cm]{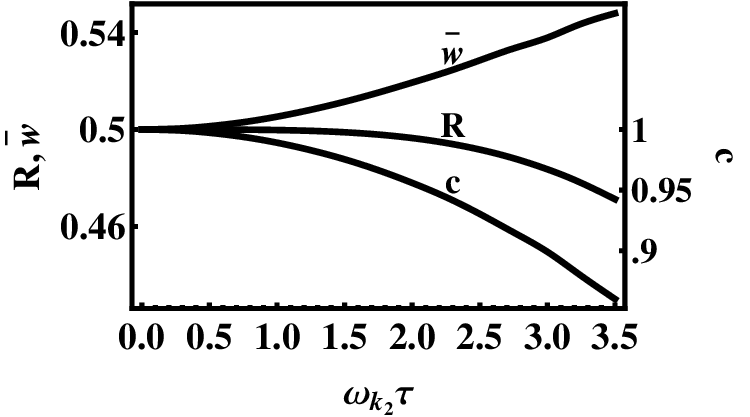}
\caption{Dependences of the maximal magnitude $R,$ background $\bar{w}$ and
contrast $c$\ on the atomic recoil frequency.}
\label{g4}
\end{figure}

\begin{figure}[!t]
\includegraphics[width=8cm]{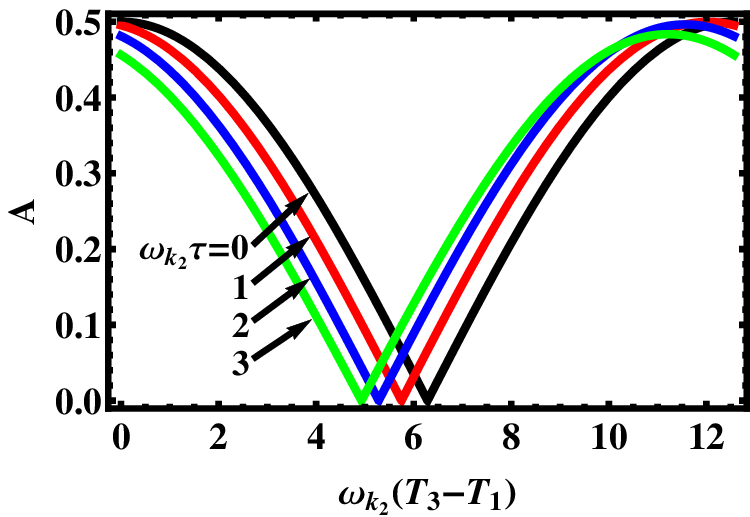}
\caption{One period of the dependence (\protect\ref{22}) on the
interrogation time $T_{3}-T_{1}$\ at the optimum conditions\ for the
different values of the recoil frequency.}
\label{f5}
\end{figure}

To maximize this magnitude one evidently has to choose $\delta _{1,\mathbf{P}%
_{0}}=\left( 1-s\right) ^{2}\omega _{k_{2}},$\ $\delta _{m}=0,$\ $\theta
_{1}=\theta _{m}=\pi /2,$\ where we defined pulses' areas as%
\begin{equation}
\theta _{n}=\Omega _{n}\tau .  \label{24}
\end{equation}%
For this choice one finds that $\left\vert x\right\vert =1/4.$\ I assumed
that 2nd and 3rd pulses still have areas $\theta _{2}=\theta _{3}=\pi /2$\
and found numerically optimum values of the Raman detunings for $s=1/2$.
These values are shown in Fig. \ref{f4}, while the maximal magnitude $R,$
background $\bar{w}$ and contrast $c=R/w$ are shown in Fig. \ref{g4}. One
period of the dependence (\ref{22}) for the different values of recoil
frequency is shown in Fig. \ref{f5}.

One sees that outside of the Raman-Nath approximation, one can achieve
almost the same magnitude of the AMZAI with the proper choice of the Raman
detunings, while the interference picture shifts as a whole. The AMZAI
becomes sensitive to the fields' frequencies, but requirements for the
frequencies' stabilization are less stringent compared to the conjugate
interferometers technique \cite{c14,c3.1} in a parameter 
\begin{equation}
\tau /T.  \label{25}
\end{equation}

In this article I consider only the Raman pulse beam splitter. I expect that
AMZAI should also be built using a standing wave beam splitter \cite{c1} and
Raman standing wave beam splitter \cite{c12}. I plan to examine these cases
elsewhere.

AMZAI can be used to measure recoil frequency $\omega _{k}.$ In contrast to
other interferometers used for these measurements \cite{c17,c20,c9,c19,c14},
AMZAI does not use counterpropagating effective wave vectors or standing
waves. Another useful feature\ is that after monitoring the response as a
function of the Raman field phase and measuring the difference between
maximal and minimal values of the response, one gets the signal (\ref{15})
insensitive to the gravity, vibration and phase noise. Calculations showed
that the similar signal occurs in the interferometers \cite{c9,c19,c14}, but
to our knowledge monitoring of the response for given time delay between
pulses and measuring the difference between response's maximum and minimum
has never been used in those experiments. Here I propose this technique to
get\ the signal sensitive only to the recoil frequency.

In this article a technique to obtain a signal insensitive to gravity is
proposed. It was justified above only for uniform gravity. I verified that
this signal can also be obtained in the presence of the gravity-gradient and
Coriolis forces, if these forces are small. Only if these forces are
sufficiently large, or the time delay between pulses is sufficiently long,
then one can use exact expressions for the atomic trajectories (\ref{10e}-%
\ref{10g}) derived in \cite{c34} to study corresponding systematic errors in
the measurement of the recoil frequency using AMZAI or technique \cite{c14}.

\acknowledgments

I appreciate fruitful discussions and suggestions from\ Paul R. Berman and
Mark A. Kasevich. This research was sponsored by Stanford University, VPFF
program, and by the Army Research Laboratory and was accomplished under
Cooperative Agreement Number W911NF-16-2-0146.

\onecolumngrid
\newpage

\appendix

\begin{center}
{\Large \textbf{Supplemental Material }}
\end{center}

\section{\label{a}MZAI with nonequal effective wave vectors}

In this section, we calculate the phase of the MZAI shown in the Fig. \ref%
{f1} Let us consider an interaction of the atom with 3 Raman pulses having
effective wave vectors $\mathbf{k}_{1}\neq \mathbf{k}_{2}\neq \mathbf{k}_{3}.
$ The choice of the wave vectors $\mathbf{k}_{i}$ and times $T_{i}$\ has to
obey 2 constraints again. From the requirement that the blue and black lines
in each port in Fig. \ref{f1} have to be parallel, i.e.%
\begin{equation}
\mathbf{k}_{1}-2\mathbf{k}_{2}+\mathbf{k}_{3}=0,  \label{m1}
\end{equation}%
The second constraint is that $D$\ is the point of crossing and therefore $%
\overrightarrow{AB}+\overrightarrow{BD}=\overrightarrow{AC}+\overrightarrow{%
CD},$\ which means that%
\begin{equation}
\mathbf{k}_{1}\left( T_{2}-T_{1}\right) +\left( \mathbf{k}_{1}-\mathbf{k}%
_{2}\right) \left( T_{3}-T_{2}\right) =\mathbf{k}_{2}\left(
T_{3}-T_{2}\right) .  \label{m2}
\end{equation}%
Resolving Eqs. (\ref{m1}, \ref{m2}) in respect to the wave vectors $\mathbf{k%
}_{1}$ and $\mathbf{k}_{3},$ one gets 
\begin{subequations}
\label{m3}
\begin{eqnarray}
\mathbf{k}_{1} &=&2\left( 1-s\right) \mathbf{k}_{2},  \label{m3a} \\
\mathbf{k}_{3} &=&2s\mathbf{k}_{2},  \label{m3b}
\end{eqnarray}%
where $s$ is still givrn by Eq. (\ref{13.1}). If initially an atomic density
matrix is given by Eq. (\ref{8}), then applying consiqently Eq. (21) from
the article \cite{c8} for the density matrix evolution before and between
the pulses and Eqs. (48) from the same article for the density matrix jump
inside given pulse, one finds that after the 3rd pulse action, the upper
state distribution is given by 
\end{subequations}
\begin{subequations}
\label{m4}
\begin{gather}
\rho _{ee}\left( \mathbf{x},\mathbf{p},T_{3+}\right) =\rho _{b}\left( 
\mathbf{x},\mathbf{p}\right) +\rho _{R}\left( \mathbf{x},\mathbf{p}\right)
+\rho _{i}\left( \mathbf{x},\mathbf{p}\right) ,  \label{m4a} \\
\rho _{b}\left( \mathbf{x},\mathbf{p}\right) =\cos ^{2}\left( \theta
_{3}/2\right) \left\{ \cos ^{2}\left( \theta _{2}/2\right) \sin ^{2}\left(
\theta _{1}/2\right) \right.   \notag \\
\times f\left[ \mathbf{X}\left( \mathbf{x}_{1},\mathbf{p}_{1}-\hbar \mathbf{k%
}_{1},-T_{1}\right) ,\mathbf{P}\left( \mathbf{x}_{1},\mathbf{p}_{1}-\hbar 
\mathbf{k}_{1},-T_{1}\right) \right] _{\left\{ \mathbf{x}_{1},\mathbf{p}%
_{1}\right\} =\left\{ \mathbf{X}\left( \mathbf{x}_{2},\mathbf{p}%
_{2},T_{1}-T_{2}\right) ,\mathbf{P}\left( \mathbf{x}_{2},\mathbf{p}%
_{2},T_{1}-T_{2}\right) \right\} }  \notag \\
+\left. \sin ^{2}\left( \theta _{2}/2\right) \cos ^{2}\left( \theta
_{1}/2\right) f\left[ \mathbf{X}\left( \mathbf{x}_{2},\mathbf{p}_{2}-\hbar 
\mathbf{k}_{2},-T_{2}\right) ,\mathbf{P}\left( \mathbf{x}_{2},\mathbf{p}%
_{2}-\hbar \mathbf{k}_{2},-T_{2}\right) \right] \right\} _{\left\{ \mathbf{x}%
_{2},\mathbf{p}_{2}\right\} =\left\{ \mathbf{X}\left( \mathbf{x},\mathbf{p}%
,T_{2}-T_{3}\right) ,\mathbf{P}\left( \mathbf{x},\mathbf{p}%
,T_{2}-T_{3}\right) \right\} }  \notag \\
+\sin ^{2}\left( \theta _{3}/2\right) \left\{ \sin ^{2}\left( \theta
_{2}/2\right) \sin ^{2}\left( \theta _{1}/2\right) \right.   \notag \\
\times f\left[ \mathbf{X}\left( \mathbf{x}_{1},\mathbf{p}_{1}-\hbar \mathbf{k%
}_{1},-T_{1}\right) ,\mathbf{P}\left( \mathbf{x}_{1},\mathbf{p}_{1}-\hbar 
\mathbf{k}_{1},-T_{1}\right) \right] _{\left\{ \mathbf{x}_{1},\mathbf{p}%
_{1}\right\} =\left\{ \mathbf{X}\left( \mathbf{x}_{2},\mathbf{p}_{2}+\hbar 
\mathbf{k}_{2},T_{1}-T_{2}\right) ,\mathbf{P}\left( \mathbf{x}_{2},\mathbf{p}%
_{2}+\hbar \mathbf{k}_{2},T_{1}-T_{2}\right) \right\} }  \notag \\
\left. +\cos ^{2}\left( \theta _{2}/2\right) \cos ^{2}\left( \theta
_{1}/2\right) f\left[ \mathbf{X}\left( \mathbf{x}_{2},\mathbf{p}%
_{2},-T_{2}\right) ,\mathbf{P}\left( \mathbf{x}_{2},\mathbf{p}%
_{2},-T_{2}\right) \right] \right\} _{\left\{ \mathbf{x}_{2},\mathbf{p}%
_{2}\right\} =\left\{ \mathbf{X}\left( \mathbf{x},\mathbf{p}-\hbar \mathbf{k}%
_{3},T_{2}-T_{3}\right) ,\mathbf{P}\left( \mathbf{x},\mathbf{p}-\hbar 
\mathbf{k}_{3},T_{2}-T_{3}\right) \right\} },  \label{m4b} \\
\rho _{R}\left( \mathbf{x},\mathbf{p}\right) =\dfrac{1}{2}\sin \theta
_{2}\sin \theta _{1}\cos \left[ \mathbf{k}_{1}\cdot \mathbf{x}_{1}-\delta
_{1}T_{1}-\phi _{1}-\left( \mathbf{k}_{2}\cdot \mathbf{x}_{2}-\delta
_{2}T_{2}-\phi _{2}\right) \right]   \notag \\
\left\{ \cos ^{2}\left( \theta _{3}/2\right) \right.   \notag \\
\times \left. f\left( 
\begin{array}{c}
\mathbf{X}\left( \mathbf{x}_{1},\mathbf{p}_{1}-\hbar \mathbf{k}%
_{1}/2,-T_{1}\right) , \\ 
\mathbf{P}\left( \mathbf{x}_{1},\mathbf{p}_{1}-\hbar \mathbf{k}%
_{1}/2,-T_{1}\right) 
\end{array}%
\right) _{\left\{ \mathbf{x}_{1},\mathbf{p}_{1}\right\} =\left\{ \mathbf{X}%
\left( \mathbf{x}_{2},\mathbf{p}_{2}-\hbar \mathbf{k}_{2}/2,T_{1}-T_{2}%
\right) ,\mathbf{P}\left( \mathbf{x}_{2},\mathbf{p}_{2}-\hbar \mathbf{k}%
_{2}/2,T_{1}-T_{2}\right) \right\} }\right\vert _{\left\{ \mathbf{x}_{2},%
\mathbf{p}_{2}\right\} =\left\{ \mathbf{X}\left( \mathbf{x},\mathbf{p}%
,T_{2}-T_{3}\right) ,\mathbf{P}\left( \mathbf{x},\mathbf{p}%
,T_{2}-T_{3}\right) \right\} }  \notag \\
-\sin ^{2}\left( \theta _{3}/2\right)   \notag \\
\times \left. \left. f\left( 
\begin{array}{c}
\mathbf{X}\left( \mathbf{x}_{1},\mathbf{p}_{1}-\hbar \mathbf{k}%
_{1}/2,-T_{1}\right) , \\ 
\mathbf{P}\left( \mathbf{x}_{1},\mathbf{p}_{1}-\hbar \mathbf{k}%
_{1}/2,-T_{1}\right) 
\end{array}%
\right) _{\left\{ \mathbf{x}_{1},\mathbf{p}_{1}\right\} =\left\{ 
\begin{array}{c}
\mathbf{X}\left( \mathbf{x}_{2},\mathbf{p}_{2}+\hbar \mathbf{k}%
_{2}/2,T_{1}-T_{2}\right) , \\ 
\mathbf{P}\left( \mathbf{x}_{2},\mathbf{p}_{2}+\hbar \mathbf{k}%
_{2}/2,T_{1}-T_{2}\right) 
\end{array}%
\right\} }\right\vert _{\left\{ \mathbf{x}_{2},\mathbf{p}_{2}\right\}
=\left\{ \mathbf{X}\left( \mathbf{x},\mathbf{p}-\hbar \mathbf{k}%
_{3},T_{2}-T_{3}\right) ,\mathbf{P}\left( \mathbf{x},\mathbf{p}-\hbar 
\mathbf{k}_{3},T_{2}-T_{3}\right) \right\} }\right\}   \notag \\
-\dfrac{1}{2}\sin \theta _{3}\left\{ \sin \theta _{2}\cos \left[ \mathbf{k}%
_{2}\cdot \mathbf{x}_{2}-\delta _{2}T_{2}-\phi _{2}-\left( \mathbf{k}%
_{3}\cdot \mathbf{x}-\delta _{3}T_{3}-\phi _{3}\right) \right] \left[ \sin
^{2}\left( \theta _{1}/2\right) \right. \right.   \notag \\
\times f\left( \mathbf{X}\left( \mathbf{x}_{1},\mathbf{p}_{1}-\hbar \mathbf{k%
}_{1},-T_{1}\right) ,\mathbf{P}\left( \mathbf{x}_{1},\mathbf{p}_{1}-\hbar 
\mathbf{k}_{1},-T_{1}\right) \right) _{\left\{ \mathbf{x}_{1},\mathbf{p}%
_{1}\right\} =\left\{ \mathbf{X}\left( \mathbf{x}_{2},\mathbf{p}_{2}+\hbar 
\mathbf{k}_{2}/2,T_{1}-T_{2}\right) ,\mathbf{P}\left( \mathbf{x}_{2},\mathbf{%
p}_{2}+\hbar \mathbf{k}_{2}/2,T_{1}-T_{2}\right) \right\} }  \notag \\
\left. -\cos ^{2}\left( \theta _{1}/2\right) f\left( \mathbf{X}\left( 
\mathbf{x}_{2},\mathbf{p}_{2}-\hbar \mathbf{k}_{2}/2,-T_{2}\right) ,\mathbf{P%
}\left( \mathbf{x}_{2},\mathbf{p}_{2}-\hbar \mathbf{k}_{2}/2,-T_{2}\right)
\right) \right]   \notag \\
-\cos ^{2}\left( \theta _{2}/2\right) \sin \theta _{1}\cos \left[ \mathbf{k}%
_{1}\mathbf{x}_{1}-\delta _{1}T_{1}-\phi _{1}-\left( \mathbf{k}_{3}\mathbf{x}%
-\delta _{3}T_{3}-\phi _{3}\right) \right]   \notag \\
\times \left. f\left[ 
\begin{array}{c}
\mathbf{X}\left( \mathbf{x}_{1},\mathbf{p}_{1}-\hbar \mathbf{k}%
_{1}/2,-T_{1}\right) , \\ 
\mathbf{P}\left( \mathbf{x}_{1},\mathbf{p}_{1}-\hbar \mathbf{k}%
_{1}/2,-T_{1}\right) 
\end{array}%
\right] _{\left\{ \mathbf{x}_{1},\mathbf{p}_{1}\right\} =\left\{ \mathbf{X}%
\left( \mathbf{x}_{2},\mathbf{p}_{2},T_{1}-T_{2}\right) ,\mathbf{P}\left( 
\mathbf{x}_{2},\mathbf{p}_{2},T_{1}-T_{2}\right) \right\} }\right\}
_{\left\{ \mathbf{x}_{2},\mathbf{p}_{2}\right\} =\left\{ \mathbf{X}\left( 
\mathbf{x},\mathbf{p}-\hbar \mathbf{k}_{3}/2,T_{2}-T_{3}\right) ,\mathbf{P}%
\left( \mathbf{x},\mathbf{p}-\hbar \mathbf{k}_{3}/2,T_{2}-T_{3}\right)
\right\} },  \label{m4c} \\
\rho _{i}\left( \mathbf{x},\mathbf{p}\right) =-\dfrac{1}{2}\sin \theta
_{3}\sin ^{2}\left( \theta _{2}/2\right) \sin \theta _{1}\cos \left[ 2\left( 
\mathbf{k}_{2}\cdot \mathbf{x}_{2}-\delta _{2}T_{2}-\phi _{2}\right) -\left( 
\mathbf{k}_{1}\cdot \mathbf{x}_{1}-\delta _{1}T_{1}-\phi _{1}\right) -\left( 
\mathbf{k}_{3}\cdot \mathbf{x}-\delta _{3}T_{3}-\phi _{3}\right) \right]  
\notag \\
\times \left. f\left[ 
\begin{array}{c}
\mathbf{X}\left( \mathbf{x}_{1},\mathbf{p}_{1}-\hbar \mathbf{k}%
_{1}/2,-T_{1}\right) , \\ 
\mathbf{P}\left( \mathbf{x}_{1},\mathbf{p}_{1}-\hbar \mathbf{k}%
_{1}/2,-T_{1}\right) 
\end{array}%
\right] _{\left\{ \mathbf{x}_{1},\mathbf{p}_{1}\right\} =\left\{ \mathbf{X}%
\left( \mathbf{x}_{2},\mathbf{p}_{2},T_{1}-T_{2}\right) ,\mathbf{P}\left( 
\mathbf{x}_{2},\mathbf{p}_{2},T_{1}-T_{2}\right) \right\} }\right\vert
_{\left\{ \mathbf{x}_{2},\mathbf{p}_{2}\right\} =\left\{ \mathbf{X}\left( 
\mathbf{x},\mathbf{p}-\hbar \mathbf{k}_{3}/2,T_{2}-T_{3}\right) ,\mathbf{P}%
\left( \mathbf{x},\mathbf{p}-\hbar \mathbf{k}_{3}/2,T_{2}-T_{3}\right)
\right\} },  \label{m4d}
\end{gather}%
where $\theta _{i}$ is the area of the pulse $i,$ $\left\{ \mathbf{X}\left( 
\mathbf{x},\mathbf{p},t\right) ,\mathbf{P}\left( \mathbf{x},\mathbf{p}%
,t\right) \right\} $ are atomic classical position and momentum subject to
initial value $\left\{ \mathbf{x},\mathbf{p}\right\} .$ The expression (\ref%
{m4}) can be used to calculate any responce associated with atoms on the
upper level. The term (\ref{m4b}) is caused by transfering of the atomic
populations between Raman pulses first considered in \cite{c35}. This term
is resposible for the background of the atomic excitation to the state $%
\left\vert e\right\rangle $.

The term (\ref{m4c}) is caused by transferring of the atomic coherence
between two Raman pulses, saturated in the field of third Raman pulse. This
term is resposible for the set of Ramsey fringes \cite{c36}. In the Doppler
limitting case 
\end{subequations}
\begin{equation}
k_{i}\dfrac{\delta p}{M}\min \left( T_{2}-T_{1},T_{3}-T_{2}\right) \gg 1,
\label{m5}
\end{equation}%
where\ $\delta p$ is the width of the atomic momentum distribution, Ramsey
fringes are washed out \cite{c37}.

The term (\ref{m4d}) is caused by the transfering coherence $\rho _{ge}$
between first and second pulse, mirroring it to the conerence $\rho _{eg}$
in the field of the second Raman pulse and further transfering between
second and third Raman pulses. This process is responsible for the atom
interference.

We use Eqs. (\ref{m4}) to get the probability of the excitation 
\begin{equation}
w=\int d\mathbf{x}d\mathbf{p}\rho _{ee}\left( \mathbf{x},\mathbf{p}%
,T_{3+}\right) .  \label{m6}
\end{equation}%
which consists of the backgrond and interference terms,%
\begin{equation}
w=\bar{w}+\tilde{w}.  \label{m7}
\end{equation}%
For the background term one finds from the Eq. (\ref{m4b})%
\begin{equation}
\bar{w}=1-\cos \theta _{3}\cos \theta _{2}\cos \theta _{1}.  \label{m8}
\end{equation}%
To get interference term from Eq. (\ref{m4d}) it is convinient to use
initial atomic position and momentum,%
\begin{equation}
\left\{ \mathbf{x}^{\prime },\mathbf{p}^{\prime }\right\} =\left\{ \mathbf{X}%
\left( \mathbf{x}_{1},\mathbf{p}_{1}-\hbar \mathbf{k}_{1}/2,-T_{1}\right) ,%
\mathbf{P}\left( \mathbf{x}_{1},\mathbf{p}_{1}-\hbar \mathbf{k}%
_{1}/2,-T_{1}\right) \right\} ,  \label{m9}
\end{equation}%
as the integration variables. Expressing other phase points in Eq. (\ref{m4d}%
) through the point (\ref{m9}) after replacement $\left\{ \mathbf{x}^{\prime
},\mathbf{p}^{\prime }\right\} \rightarrow \left\{ \mathbf{x},\mathbf{p}%
\right\} $ one gets
\begin{subequations}
\label{m10}
\begin{gather}
\tilde{w}=-\dfrac{1}{2}\sin \theta _{3}\sin ^{2}\left( \theta _{2}/2\right)
\sin \theta _{1}\int d\mathbf{x}d\mathbf{p}f\left( \mathbf{x},\mathbf{p}%
\right) \cos \left[ \phi \left( \mathbf{x},\mathbf{p}\right) -\delta
_{1}T_{1}+2\delta _{2}T_{2}-\delta _{3}T_{3}-\phi _{1}+2\phi _{2}-\phi _{3}%
\right] ,  \label{m10a} \\
\phi \left( \mathbf{x},\mathbf{p}\right) =\mathbf{k}_{1}\cdot \mathbf{x}%
_{1}-2\mathbf{k}_{2}\cdot \mathbf{x}_{2}+\mathbf{k}_{3}\cdot \mathbf{x}_{3},
\label{m10b} \\
\left\{ \mathbf{x}_{1},\mathbf{p}_{1}\right\} =\left\{ \mathbf{X}\left( 
\mathbf{x},\mathbf{p},T_{1}\right) ,\mathbf{P}\left( \mathbf{x},\mathbf{p}%
,T_{1}\right) +\hbar \mathbf{k}_{1}/2\right\} ,  \label{m10c} \\
\left\{ \mathbf{x}_{2},\mathbf{p}_{2}\right\} =\left\{ \mathbf{X}\left( 
\mathbf{x}_{1},\mathbf{p}_{1},T_{2}-T_{1}\right) ,\mathbf{P}\left( \mathbf{x}%
_{1},\mathbf{p}_{1},T_{2}-T_{1}\right) \right\} ,  \label{m10d} \\
\left\{ \mathbf{x}_{3},\mathbf{p}_{3}\right\} =\left\{ \mathbf{X}\left( 
\mathbf{x}_{2},\mathbf{p}_{2},T_{3}-T_{2}\right) ,\mathbf{P}\left( \mathbf{x}%
_{2},\mathbf{p}_{2},T_{3}-T_{2}\right) +\hbar \mathbf{k}_{3}/2\right\} ,
\label{m10e}
\end{gather}%
where the phase of the MZAI is given by Eq. (\ref{m10b}). In the case of the
uniform gravity field $\mathbf{g}$%
\end{subequations}
\begin{equation}
\left\{ \mathbf{X}\left( \mathbf{x},\mathbf{p},t\right) ,\mathbf{P}\left( 
\mathbf{x},\mathbf{p},t\right) \right\} =\left\{ \mathbf{x}+\dfrac{\mathbf{p}%
}{M}t+\dfrac{\mathbf{g}t^{2}}{2},\mathbf{p}+M\mathbf{g}t\right\} ,
\label{m11}
\end{equation}%
for the wave vectors (\ref{m3}) one arrives for the phase%
\begin{equation}
\phi \left( \mathbf{x},\mathbf{p}\right) =s\left( 1-s\right) \mathbf{k}%
_{2}\cdot \mathbf{g}\left( T_{3}-T_{1}\right) ^{2},  \label{m12}
\end{equation}%
which still contains no quantum term.

\section{\label{b}Atomic amplitudes in the momentum state}

\QTP{Body Math}
In this appendix, we use the expressions for the atomic amplitudes'
evolution in the uniform field \cite{c11}. For the purpose of integrity we
consider both the amplitudes' evolution between Raman pulses \cite{c11} and
during the interaction with a given pulse. Consider an interaction of the
3-level atom with a rectangular pulse of the 2 traveling waves 
\begin{equation}
\mathbf{E}\left( \mathbf{x},t\right) =\left\{ \mathbf{E}_{1}\exp \left[
i\left( \mathbf{q}_{1}\cdot \mathbf{x}-\omega _{1}t-\phi ^{\left( 1\right)
}\left( t\right) \right) \right] +\mathbf{E}_{2}\exp \left[ i\left( \mathbf{q%
}_{2}\cdot \mathbf{x}-\omega _{2}t-\phi ^{\left( 2\right) }\left( t\right)
\right) \right] \right\} f\left( t\right) +c.c.  \label{a1}
\end{equation}%
where $\mathbf{E}_{i},$ $\mathbf{q}_{i},$ $\omega _{i}$ and $\phi ^{\left(
i\right) }\left( t\right) $ are amplitude, wave vector, frequency and phase
of the waves,%
\begin{equation}
f\left( t\right) =\left\{ 
\begin{array}{c}
1,\text{ for }T<t<T+\tau \\ 
0,\text{ for }t<T\text{ and }t>T+\tau%
\end{array}%
\right.  \label{a2}
\end{equation}%
is the shape of the rectangular pulse acting at the moment $T$ and having a
duration $\tau .$ We assume that the 2 atomic states, $g$ and $e,$ are the
sublevels of the atomic ground state, while the third state $0$ is an
excited state, fields $\mathbf{E}_{1}$ and $\mathbf{E}_{2}$ are resonant to
the transitions $g\rightarrow 0$ and $e\rightarrow 0$ correspondingly. The
Hamiltonian of the atom is given by%
\begin{equation}
H=\dfrac{\mathbf{p}^{2}}{2M}-M\mathbf{g}\cdot \mathbf{x}+\hbar \left\{
\Omega ^{\left( 1\right) }\exp \left[ i\left( \mathbf{q}_{1}\cdot \mathbf{x}%
-\Delta _{1}t-\phi ^{\left( 1\right) }\left( t\right) \right) \right]
\left\vert 0\right\rangle \left\langle g\right\vert +\Omega ^{\left(
2\right) }\exp \left[ i\left( \mathbf{q}_{2}\cdot \mathbf{x}-\Delta
_{2}t-\phi ^{\left( 2\right) }\left( t\right) \right) \right] \left\vert
0\right\rangle \left\langle e\right\vert +H.c.\right\} ,  \label{a3}
\end{equation}%
where $\Omega ^{\left( 1\right) }\equiv -\mathbf{d}_{0g}\cdot \mathbf{E}%
_{1}/\hbar $ and $\Omega ^{\left( 2\right) }\equiv -\mathbf{d}_{0e}\cdot 
\mathbf{E}_{2}/\hbar $ are Rabi frequencies, $\mathbf{d}_{0g}$ and $\mathbf{d%
}_{0e}$ are the matrix elements of the dipole moment operator, $\Delta
_{1}=\omega _{1}-\omega _{0g}\ $and $\Delta _{2}=\omega _{2}-\omega _{0e}$
are the detunings of the frequencies. Let us \ eliminate an amplitude of the
excited state, which evolves as 
\begin{eqnarray}
i\dot{a}\left( 0,\mathbf{p},t\right) &=&\left( \dfrac{\mathbf{p}^{2}}{%
2M\hbar }-iM\mathbf{g}\cdot \partial _{\mathbf{p}}\right) a\left( 0,\mathbf{p%
},t\right)  \notag \\
&&+\left\{ \Omega ^{\left( 1\right) }e^{-i\left[ \Delta _{1}t+\phi ^{\left(
1\right) }\left( t\right) \right] }a\left( g,\mathbf{p}-\hbar \mathbf{q}%
_{1},t\right) +\Omega ^{\left( 2\right) }e^{-i\left[ \Delta _{2}t+\phi
^{\left( 2\right) }\left( t\right) \right] }a\left( e,\mathbf{p}-\hbar 
\mathbf{q}_{2},t\right) \right\} f\left( t\right) .  \label{a4}
\end{eqnarray}%
When the detunings are closed 
\begin{equation}
\Delta _{1}\approx \Delta _{2}\approx \Delta  \label{a5}
\end{equation}%
and sufficiently large, 
\begin{equation}
\left\vert \Delta \right\vert \gg Max\left\{ \left\vert \tilde{\delta}%
\right\vert ,\tau ^{-1},\left\vert \dot{\phi}^{\left( i\right) }\right\vert
,q_{i}gT,q_{i}\dfrac{p}{M},\omega _{q_{i}}\right\} ,  \label{a6}
\end{equation}%
where 
\begin{equation}
\tilde{\delta}\equiv \Delta _{1}-\Delta _{2}  \label{a7}
\end{equation}%
is Raman detuning, one can neglect the first term in the right-hand-side of
Eq. (\ref{a4}) to get 
\begin{equation}
a\left( 0,\mathbf{p},t\right) =\left\{ \Omega ^{\left( 1\right) }\exp \left[
-i\left( \Delta _{1}t+\phi ^{\left( 1\right) }\left( t\right) \right) \right]
a\left( g,\mathbf{p}-\hbar \mathbf{q}_{1},t\right) +\Omega ^{\left( 2\right)
}\exp \left[ -i\left( \Delta _{2}t+\phi ^{\left( 2\right) }\left( t\right)
\right) \right] a\left( e,\mathbf{p}-\hbar \mathbf{q}_{2},t\right) \right\}
/\Delta ,  \label{a8}
\end{equation}%
where the usual assumption was accepted that the atomic ground state
amplitudes vary slowly at the time $\left\vert \Delta \right\vert ^{-1}.$
Substituting this expression in the equations for the atomic ground state
amplitudes, one gets 
\begin{subequations}
\label{a9}
\begin{eqnarray}
i\left( \partial _{t}+M\mathbf{g}\cdot \partial _{\mathbf{p}}+i\dfrac{p^{2}}{%
2M\hbar }\right) \tilde{a}\left( e,\mathbf{p},t\right) &=&\dfrac{\Omega }{2}%
f\left( t\right) \exp \left[ -i\delta t-i\phi \left( t\right) \right] \tilde{%
a}\left( g,\mathbf{p}-\hbar \mathbf{k},t\right) ,  \label{a9a} \\
i\left( \partial _{t}+M\mathbf{g}\cdot \partial _{\mathbf{p}}+i\dfrac{p^{2}}{%
2M\hbar }\right) \tilde{a}\left( g,\mathbf{p},t\right) &=&\dfrac{\Omega
^{\ast }}{2}f\left( t\right) \exp \left[ i\delta t+i\phi \left( t\right) %
\right] \tilde{a}\left( e,\mathbf{p}+\hbar \mathbf{k},t\right) ,  \label{a9b}
\end{eqnarray}%
where $\tilde{a}\left( g,\mathbf{p},t\right) =\exp \left( i\left\vert \Omega
^{\left( 1\right) }\right\vert ^{2}t/\Delta \right) a\left( g,\mathbf{p}%
,t\right) ;\tilde{a}\left( e,\mathbf{p},t\right) =\exp \left( i\left\vert
\Omega ^{\left( 2\right) }\right\vert ^{2}t/\Delta \right) a\left( g,\mathbf{%
p},t\right) ,$ $\mathbf{k}=\mathbf{q}_{1}-\mathbf{q}_{2}$ is an effective
wave vector, $\delta =\tilde{\delta}-\left( \left\vert \Omega ^{\left(
2\right) }\right\vert ^{2}-\left\vert \Omega ^{\left( 1\right) }\right\vert
^{2}\right) /\Delta ,$ $\Omega =2\Omega ^{\left( 1\right) }\Omega ^{\left(
2\right) \ast }/\Delta $ is Rabi Raman frequency, $\phi \left( t\right)
=\phi ^{\left( 1\right) }\left( t\right) -\phi ^{\left( 2\right) }\left(
t\right) .$ In the accelerated frame 
\end{subequations}
\begin{equation}
\mathbf{p}=\mathbf{P}+M\mathbf{g}t,  \label{a10}
\end{equation}%
using the initial atomic momentum $\mathbf{P}$ as an independent variable,
one obtains 
\begin{subequations}
\label{a11}
\begin{eqnarray}
i\left( \partial _{t}+i\dfrac{\left( \mathbf{P}+M\mathbf{g}t\right) ^{2}}{%
2M\hbar }\right) \tilde{a}\left( e,\mathbf{P},t\right) &=&\dfrac{\Omega }{2}%
f\left( t\right) \exp \left[ -i\delta t-i\phi \left( t\right) \right] \tilde{%
a}\left( g,\mathbf{P}-\hbar \mathbf{k},t\right) ,  \label{a11a} \\
i\left( \partial _{t}+i\dfrac{\left( \mathbf{P}+M\mathbf{g}t\right) ^{2}}{%
2M\hbar }\right) \tilde{a}\left( g,\mathbf{P},t\right) &=&\dfrac{\Omega
^{\ast }}{2}f\left( t\right) \exp \left[ i\delta t+i\phi \left( t\right) %
\right] \tilde{a}\left( e,\mathbf{P}+\hbar \mathbf{k},t\right) .
\label{a11b}
\end{eqnarray}%
Then one finds that in the interaction representation 
\end{subequations}
\begin{equation}
\tilde{a}\left( n,\mathbf{P},t\right) =\exp \left[ -i\int_{0}^{t}dt^{\prime }%
\dfrac{\left( \mathbf{P}+M\mathbf{g}t^{\prime }\right) ^{2}}{2M\hbar }\right]
c\left( n,\mathbf{P},t\right)  \label{a12}
\end{equation}%
the vector state%
\begin{equation}
c\left( t\right) =\left( 
\begin{array}{c}
c\left( e,\mathbf{P}+\dfrac{\hbar \mathbf{k}}{2},t\right) \\ 
c\left( g,\mathbf{P}-\dfrac{\hbar \mathbf{k}}{2},t\right)%
\end{array}%
\right)  \label{a13}
\end{equation}%
evolves as%
\begin{equation}
i\dot{c}=\dfrac{f\left( t\right) }{2}\left( 
\begin{array}{cc}
0 & \Omega \exp \left\{ -i\left[ \delta t+\phi \left( t\right) -\mathbf{%
k\cdot }\dfrac{\mathbf{P}}{M}t-\dfrac{1}{2}\mathbf{k\cdot g}t^{2}\right]
\right\} \\ 
\Omega ^{\ast }\exp \left\{ i\left[ \delta t+\phi \left( t\right) -\mathbf{%
k\cdot }\dfrac{\mathbf{P}}{M}t-\dfrac{1}{2}\mathbf{k\cdot g}t^{2}\right]
\right\} & 0%
\end{array}%
\right) c.  \label{a14}
\end{equation}%
One sees that in the accelerated frame the atomic amplitude in the
interaction representation (\ref{a12}) stays unchanged at the pulses' free
time,%
\begin{equation}
c\left( t\right) =\text{constant, at }t<T\text{ or }t>T+\tau .  \label{a14.1}
\end{equation}

\QTP{Body Math}
If one chirps linearly the fields' frequencies, i.e. if%
\begin{equation}
\phi \left( t\right) =\phi +\alpha t^{2}/2  \label{a15}
\end{equation}%
and if the chirping rate $\alpha $ is close to $\mathbf{k\cdot g},$ 
\begin{equation}
\left\vert \alpha -\mathbf{k\cdot g}\right\vert \tau ^{2}\ll 1,  \label{a16}
\end{equation}%
then one can neglect the quadratic over%
\begin{equation}
t^{\prime }=t-T  \label{a17}
\end{equation}%
terms in the phase factors in Eq. (\ref{a14}) and arrives at the well-known
equation for the amplitudes of a two-level atom in the rectangular resonant
pulse with permanent detuning and phase,%
\begin{equation}
i\dfrac{dc}{dt^{\prime }}=\dfrac{f\left( T+t^{\prime }\right) }{2}\left( 
\begin{array}{cc}
0 & \Omega \exp \left\{ -i\left[ \delta _{\mathbf{P}}t^{\prime }+\phi \left( 
\mathbf{P}\right) \right] \right\} \\ 
\Omega ^{\ast }\exp \left\{ i\left[ \delta _{\mathbf{P}}t^{\prime }+\phi
\left( \mathbf{P}\right) \right] \right\} & 0%
\end{array}%
\right) c,  \label{a18}
\end{equation}%
where 
\begin{subequations}
\begin{eqnarray}
\phi \left( \mathbf{P}\right) &=&\phi +\left( \delta -\mathbf{k\cdot }\dfrac{%
\mathbf{P}}{M}\right) T+\dfrac{1}{2}\left( \alpha -\mathbf{k\cdot g}\right)
T^{2},  \label{a19a} \\
\delta _{\mathbf{P}} &=&\partial \phi \left( \mathbf{P}\right) /\partial T.
\label{a19b}
\end{eqnarray}%
Using the solution of the Eq. (\ref{a18}), one finds that during an
interaction with Raman pulse the atomic amplitudes jump as 
\end{subequations}
\begin{subequations}
\label{a20}
\begin{eqnarray}
c\left( e,\mathbf{P},T+\tau \right) &=&F_{ee}\left( \mathbf{P}-\hbar \mathbf{%
k}/2\right) c\left( e,\mathbf{P},T\right) +F_{eg}\left( \mathbf{P}-\hbar 
\mathbf{k}/2\right) c\left( g,\mathbf{P}-\hbar \mathbf{k},T\right) ,
\label{a20a} \\
c\left( g,\mathbf{P},T+\tau \right) &=&F_{ge}\left( \mathbf{P}+\hbar \mathbf{%
k}/2\right) c\left( e,\mathbf{P}+\hbar \mathbf{k},T\right) +F_{gg}\left( 
\mathbf{P}+\hbar \mathbf{k}/2\right) c\left( g,\mathbf{P},T\right) ,
\label{a20b}
\end{eqnarray}%
where $F_{m,n}\left( \mathbf{P}\right) $ is an element of the matrix 
\end{subequations}
\begin{equation}
F\left( \mathbf{P}\right) =\left( 
\begin{array}{cc}
f_{ee}\left( \mathbf{P}\right) & f_{eg}\left( \mathbf{P}\right) \exp \left\{
-i\phi \left( \mathbf{P}\right) \right\} \\ 
f_{ge}\left( \mathbf{P}\right) \exp \left\{ i\phi \left( \mathbf{P}\right)
\right\} & f_{gg}\left( \mathbf{P}\right)%
\end{array}%
\right) ,  \label{a21}
\end{equation}%
where 
\begin{subequations}
\label{a22}
\begin{eqnarray}
f\left( \mathbf{P}\right) &=&\left( 
\begin{array}{cc}
\exp \left\{ -i\delta _{\mathbf{P}}\tau /2\right\} f_{d}\left[ \Omega
,\delta _{\mathbf{P}}\right] & -i\exp \left\{ -i\delta _{\mathbf{P}}\tau
/2\right\} f_{a}\left[ \Omega ,\delta _{\mathbf{P}}\right] \\ 
-i\exp \left\{ i\delta _{\mathbf{P}}\tau /2\right\} f_{a}^{\ast }\left[
\Omega ,\delta _{\mathbf{P}}\right] & \exp \left\{ i\delta _{\mathbf{P}}\tau
/2\right\} f_{d}^{\ast }\left[ \Omega ,\delta _{\mathbf{P}}\right]%
\end{array}%
\right) ,  \label{a22a} \\
f_{d}\left( \Omega ,\delta \right) &=&\cos \dfrac{\Omega _{r}\left( \Omega
,\delta \right) \tau }{2}+i\dfrac{\delta }{\Omega _{r}\left( \Omega ,\delta
\right) }\sin \dfrac{\Omega _{r}\left( \Omega ,\delta \right) \tau }{2},
\label{a22b} \\
f_{a}\left( \Omega ,\delta \right) &=&\dfrac{\Omega }{\Omega _{r}\left(
\Omega ,\delta \right) }\sin \dfrac{\Omega _{r}\left( \Omega ,\delta \right)
\tau }{2},  \label{a22c} \\
\Omega _{r}\left( \Omega ,\delta \right) &=&\sqrt{\left\vert \Omega
\right\vert ^{2}+\delta ^{2}}.  \label{a22d}
\end{eqnarray}

\section{\label{c}AMZAI response}

\QTP{Body Math}
Let us now apply Eqs. (\ref{a14.1}, \ref{a20}) to consider AMZAI, i.e. to
calculate the total probability of the atoms' excitation 
\end{subequations}
\begin{equation}
w=\int d\mathbf{P}\left\vert c\left( e,\mathbf{P},T_{3}+\tau \right)
\right\vert ^{2}  \label{a23}
\end{equation}%
with 3 Raman pulses and one microwave pulse having Rabi frequencies $\left\{
\Omega _{1},\Omega _{2},\Omega _{m},\Omega _{3}\right\} ,$ detunings $%
\left\{ \delta _{1},\delta _{2},\delta _{m},\delta _{3}\right\} ,$ phases $%
\left\{ \phi _{1},\phi _{2},\phi _{m},\phi _{3}\right\} ,$ wave vectors and
chirping rates 
\begin{subequations}
\begin{eqnarray}
\left\{ \mathbf{k}_{1},\mathbf{k}_{2},\mathbf{k}_{m},\mathbf{k}_{3}\right\}
&=&\mathbf{k}_{2}\left\{ 1-s,1,0,s\right\} ,  \label{a24a} \\
\left\{ \alpha _{1},\alpha _{2},\alpha _{m},\alpha _{3}\right\} &=&\alpha
_{2}\left\{ 1-s,1,0,s\right\} ,  \label{a24b}
\end{eqnarray}%
and acting at the moments $\left\{ T_{1},T_{2},T_{m},T_{3}\right\} .$
Assuming that before the pulses action atoms are in the ground state only $%
\left[ c\left( e,\mathbf{P},T_{1}\right) =0\right] ,$ and using consequently
the Eqs. (\ref{a20}) for each pulse one arrives to the following result 
\end{subequations}
\begin{eqnarray}
c\left( e,\mathbf{P},T_{3}+\tau \right) &=&c_{1+}+c_{2+}+c_{1-}+c_{2-} 
\notag \\
&&+F_{3ee}\left( \mathbf{P}-s\hbar \mathbf{k}_{2}/2\right) \left[
F_{mee}F_{2ee}\left( \mathbf{P}-\hbar \mathbf{k}_{2}/2\right) F_{1eg}\left( 
\mathbf{P}-\left( 1-s\right) \hbar \mathbf{k}_{2}/2\right) c\left( g,\mathbf{%
P}-\left( 1-s\right) \hbar \mathbf{k}_{2},T_{1}\right) \right.  \notag \\
&&\left. +F_{meg}F_{2ge}\left( \mathbf{P}+\hbar \mathbf{k}_{2}/2\right)
F_{1eg}\left( \mathbf{P}+\left( 1+s\right) \hbar \mathbf{k}_{2}/2\right)
c\left( g,\mathbf{P}+s\hbar \mathbf{k}_{2},T_{1}\right) \right]  \notag \\
&&+F_{3eg}\left( \mathbf{P}-s\hbar \mathbf{k}_{2}/2\right) \left[
F_{mge}F_{2eg}\left( \mathbf{P}-\left( 2s+1\right) \hbar \mathbf{k}%
_{2}/2\right) F_{1gg}\left( \mathbf{P}-\left( 3s+1\right) \hbar \mathbf{k}%
_{2}/2\right) c\left( g,\mathbf{P}-\left( s+1\right) \hbar \mathbf{k}%
_{2},T_{1}\right) \right.  \notag \\
&&+\left. F_{mgg}F_{2gg}\left( \mathbf{P}-\left( 2s-1\right) \hbar \mathbf{k}%
_{2}/2\right) F_{1gg}\left( \mathbf{P}-\left( 3s-1\right) \hbar \mathbf{k}%
_{2}/2\right) c\left( g,\mathbf{P}-s\hbar \mathbf{k}_{2},T_{1}\right) \right]
,  \label{a25}
\end{eqnarray}%
where 
\begin{subequations}
\label{a26}
\begin{eqnarray}
c_{1+} &=&F_{3ee}\left( \mathbf{P}-s\hbar \mathbf{k}_{2}/2\right)
F_{mee}F_{2eg}\left( \mathbf{P}-\hbar \mathbf{k}_{2}/2\right) F_{1gg}\left( 
\mathbf{P}-\left( 1+s\right) \hbar \mathbf{k}_{2}/2\right) c\left( g,\mathbf{%
P}-\hbar \mathbf{k}_{2},T_{1}\right) ,  \label{a26a} \\
c_{2+} &=&F_{3eg}\left( \mathbf{P}-s\hbar \mathbf{k}_{2}/2\right)
F_{mge}F_{2ee}\left( \mathbf{P}-\left( 2s+1\right) \hbar \mathbf{k}%
_{2}/2\right) F_{1eg}\left( \mathbf{P}-\left( 1+s\right) \hbar \mathbf{k}%
_{2}/2\right) c\left( g,\mathbf{P}-\hbar \mathbf{k}_{2},T_{1}\right) ,
\label{a26b} \\
c_{1-} &=&F_{3ee}\left( \mathbf{P}-s\hbar \mathbf{k}_{2}/2\right)
F_{meg}F_{2gg}\left( \mathbf{P}+\hbar \mathbf{k}_{2}/2\right) F_{1gg}\left( 
\mathbf{P}+\left( 1-s\right) \hbar \mathbf{k}_{2}/2\right) c\left( g,\mathbf{%
P},T_{1}\right) ,  \label{a26c} \\
c_{2-} &=&F_{3eg}\left( \mathbf{P}-s\hbar \mathbf{k}_{2}/2\right)
F_{mgg}F_{2ge}\left( \mathbf{P}-\left( 2s-1\right) \hbar \mathbf{k}%
_{2}/2\right) F_{1eg}\left( \mathbf{P}+\left( 1-s\right) \hbar \mathbf{k}%
_{2}/2\right) c\left( g,\mathbf{P},T_{1}\right) ,  \label{a26d}
\end{eqnarray}%
$F_{N\alpha \beta }$ is $\left( \alpha \beta \right) -$element of matrix (%
\ref{a21}) associated with the pulse $N.$ The sum of the 8 integrals of the
absolute squares of the each term in (\ref{a25}) is a background $\bar{w}$
of the response (\ref{a23})$,$ while the cross terms are responsible for the
interference. If the initial atomic distribution $\left\vert c\left( g,%
\mathbf{P},T_{1}\right) \right\vert ^{2}$ is centered near the momentum $%
\mathbf{P}_{0}$ and, after using the filtering technique \cite{c10,c33}, it
is narrowed to the width 
\end{subequations}
\begin{equation}
\delta P\ll \dfrac{M}{k\tau }  \label{a27}
\end{equation}%
and if the pulse duration is comparable to the inverse recoil frequency,
then different terms in Eq. (\ref{a25}) are centered in the momentum space
near the points which are far distanced from one\ another and cannot
interfere. There are only 2 exclusions, 2 pairs of the terms $\left\{
c_{1+,}c_{2+}\right\} ,$ located at the point $\mathbf{P}=\mathbf{P}%
_{0}+\hbar \mathbf{k}_{2},$ and $\left\{ c_{1-,}c_{2-}\right\} ,$ located at
the point $\mathbf{P}=\mathbf{P}_{0},$ leading to the interferometric part
of the excitation probability 
\begin{subequations}
\label{a28}
\begin{eqnarray}
\tilde{w} &=&w_{+}+w_{-},  \label{a28a} \\
w_{\pm } &=&2\func{Re}\int d\mathbf{P}c_{1\pm }c_{2\pm }^{\ast }
\label{a28b}
\end{eqnarray}%
For the parameter $s$ given by Eq. (\ref{13.1}) the rapid oscillations of
the integrands in Eq. (\ref{a28b}) as functions of the momentum with period $%
\thicksim M/k_{2}T$ disappear. Other factors originated from the matrices $%
f_{N}\left( \mathbf{P}\right) ,$ given by Eq. (\ref{a22a}), are smooth
functions of the momentum with size $\thicksim M/k\tau ,$ so that one can
put in Eqs. (\ref{a26}) $c\left( g,\mathbf{P},T_{1}\right) =\sqrt{\delta
\left( \mathbf{P}-\mathbf{P}_{0}\right) }$ to find 
\end{subequations}
\begin{subequations}
\label{a29}
\begin{eqnarray}
w_{\pm } &=&-2\func{Re}\left\{ \exp \left[ i\left( -\phi _{I}-\phi
_{c}+\left( -\delta _{m}+\delta _{1,\mathbf{P}_{0}}-\omega _{k_{2}}\left(
1-s\right) ^{2}\right) \tau \right) \mp r\right] xx_{\pm }\right\} ,
\label{a29a} \\
r &=&\phi _{q}-s\omega _{k_{2}}\tau ;  \label{a29b} \\
\phi _{c} &=&\dfrac{1}{2}\left( \mathbf{k}_{2}\cdot \mathbf{g}-\alpha
_{2}\right) s\left( 1-s\right) \left( T_{3}-T_{1}\right) ^{2},  \label{a29c}
\\
x &=&f_{d}^{\ast }\left[ \Omega _{1},\delta _{1,\mathbf{P}_{0}}-\left(
1-s\right) ^{2}\omega _{k_{2}}\right] f_{a}\left[ \left\vert \Omega
_{1}\right\vert ,\delta _{1,\mathbf{P}_{0}}-\left( 1-s\right) ^{2}\omega
_{k_{2}}\right] f_{a}\left[ \left\vert \Omega _{m}\right\vert ,\delta _{m}%
\right] f_{d}\left[ \Omega _{m},\delta _{m}\right] ,  \label{a29d} \\
x_{+} &=&f_{d}\left[ \Omega _{3},\delta _{3,\mathbf{P}_{0}}-s\left(
2-s\right) \omega _{k_{2}}\right] f_{a}\left[ \left\vert \Omega
_{3}\right\vert ,\delta _{3,\mathbf{P}_{0}}-s\left( 2-s\right) \omega
_{k_{2}}\right]  \notag \\
&&\times f_{a}\left[ \left\vert \Omega _{2}\right\vert ,\delta _{2,\mathbf{P}%
_{0}}-\omega _{k_{2}}\right] f_{d}^{\ast }\left[ \Omega _{2},\delta _{2,%
\mathbf{P}_{0}}-\left( 1-2s\right) \omega _{k_{2}}\right] ,  \label{a29e} \\
x_{-} &=&f_{d}\left[ \Omega _{3},\delta _{3,\mathbf{P}_{0}}+s^{2}\omega
_{k_{2}}\right] f_{a}\left[ \left\vert \Omega _{3}\right\vert ,\delta _{3,%
\mathbf{P}_{0}}+s^{2}\omega _{k_{2}}\right]  \notag \\
&&\times f_{d}^{\ast }\left[ \Omega _{2},\delta _{2,\mathbf{P}_{0}}-\omega
_{k_{2}}\right] f_{a}\left[ \left\vert \Omega _{2}\right\vert ,\delta _{2,%
\mathbf{P}_{0}}+\left( 2s-1\right) \omega _{k_{2}}\right] ,  \label{a29f}
\end{eqnarray}%
where the quantum phase $\phi _{q}$ and phase $\phi _{I}$ are given by Eqs.(%
\ref{14c}, \ref{10c.1}), and Eq. (\ref{a29c}) now defines the classical
phase $\phi _{c}.$

\QTP{Body Math}
Calculating the absolute squares of the terms in Eq, (\ref{a25}), one
arrives at the following results for the background term 
\end{subequations}
\begin{eqnarray}
\bar{w} &=&\left\vert f_{a}\left[ \Omega _{1},\delta _{1,\mathbf{P}%
_{0}}-\left( 1-s\right) ^{2}\omega _{k_{2}}\right] \right\vert ^{2}\left\{
\left\vert f_{d}\left[ \Omega _{m},\delta _{m}\right] \right\vert ^{2}\left[
\left\vert f_{d}\left( \Omega _{3},\delta _{3,\mathbf{P}_{0}}-s\left(
2-s\right) \omega _{k_{2}}\right) f_{d}\left( \Omega _{2},\delta _{2,\mathbf{%
P}_{0}}-\left( 1-2s\right) \omega _{k_{2}}\right) \right\vert ^{2}\right.
\right.  \notag \\
&&\left. +\left\vert f_{a}\left( \Omega _{3},\delta _{3,\mathbf{P}%
_{0}}+s^{2}\omega _{k_{2}}\right) f_{a}\left( \Omega _{2},\delta _{2,\mathbf{%
P}_{0}}-\left( 1-2s\right) \omega _{k_{2}}\right) \right\vert ^{2}\right] 
\notag \\
&&+\left\vert f_{a}\left[ \Omega _{m},\delta _{m}\right] \right\vert ^{2}%
\left[ \left\vert f_{d}\left( \Omega _{3},\delta _{3,\mathbf{P}%
_{0}}+3s^{2}\omega _{k_{2}}\right) f_{a}\left( \Omega _{2},\delta _{2,%
\mathbf{P}_{0}}-\left( 1-2s\right) \omega _{k_{2}}\right) \right\vert
^{2}\right.  \notag \\
&&\left. \left. +\left\vert f_{a}\left( \Omega _{3},\delta _{3,\mathbf{P}%
_{0}}-s\left( 2-s\right) \omega _{k_{2}}\right) f_{d}\left( \Omega
_{2},\delta _{2,\mathbf{P}_{0}}-\left( 1-2s\right) \omega _{k_{2}}\right)
\right\vert ^{2}\right] \right\}  \notag \\
&&+\left\vert f_{d}\left[ \Omega _{1},\delta _{1,\mathbf{P}_{0}}-\left(
1-s\right) ^{2}\omega _{k_{2}}\right] \right\vert ^{2}\left\{ \left\vert
f_{d}\left[ \Omega _{m},\delta _{m}\right] \right\vert ^{2}\left[ \left\vert
f_{d}\left( \Omega _{3},\delta _{3,\mathbf{P}_{0}}-s\left( 2-s\right) \omega
_{k_{2}}\right) f_{a}\left( \Omega _{2},\delta _{2,\mathbf{P}_{0}}-\omega
_{k_{2}}\right) \right\vert ^{2}\right. \right.  \notag \\
&&+\left. \left\vert f_{a}\left( \Omega _{3},\delta _{3,\mathbf{P}%
_{0}}-s^{2}\omega _{k_{2}}\right) f_{d}\left( \Omega _{2},\delta _{2,\mathbf{%
P}_{0}}-\omega _{k_{2}}\right) \right\vert ^{2}\right]  \notag \\
&&+\left\vert f_{a}\left[ \Omega _{m},\delta _{m}\right] \right\vert ^{2}%
\left[ \left\vert f_{d}\left( \Omega _{3},\delta _{3,\mathbf{P}%
_{0}}+s^{2}\omega _{k_{2}}\right) f_{d}\left( \Omega _{2},\delta _{2,\mathbf{%
P}_{0}}-\omega _{k_{2}}\right) \right\vert ^{2}\right.  \notag \\
&&\left. \left. +\left\vert f_{a}\left( \Omega _{3},\delta _{3,\mathbf{P}%
_{0}}-s\left( 2+s\right) \omega _{k_{2}}\right) f_{a}\left( \Omega
_{2},\delta _{2,\mathbf{P}_{0}}-\omega _{k_{2}}\right) \right\vert ^{2}%
\right] \right\} .  \label{a30}
\end{eqnarray}


\begin{thebibliography}{99}
\bibitem{c11.1} One can expect that a necessity of the microwave pulse
disappears if one uses the standing wave beam splitter \cite{c1} in the
Raman-Nath or Bragg regime, which does not change the internal state of the
atom.
\end{thebibliography}
\end{document}